\documentclass[12pt,amsfonts]{article}    

\usepackage{amsfonts}

\def\half{\textstyle{\frac{1}{2}}}
\def\threequarter{\textstyle{\frac{3}{4}}}

\def\p{\phi}

\def\l{\lambda}

\def\ra{\rightarrow}
\def\tint{{\textstyle\int}}
\def\hg{{\hat g}}
\def\hp{{\hat\pi}}
\def\hph{{\hat\phi}}
\def\s{\hskip.08em}

\def\b{\begin{eqnarray*}}  
\def\e{\end{eqnarray*}}    
\def\bn{\begin{eqnarray}}  
\def\en{\end{eqnarray}}   

\def\<{\langle}
\def\>{\rangle}

\def\no{\nonumber}

\def\quarter{\textstyle{\frac{1}{4}}}

\def\{{\lbrace}
\def\}{\rbrace}
\def\pp{\partial}
\begin{document}

\title{The Benefits of Affine Quantization}        
\author{John R. Klauder\footnote{klauder@phys.ufl.edu} \\
Department of Physics and Department of Mathematics \\
University of Florida,   
Gainesville, FL 32611-8440}
\date{ }
\bibliographystyle{unsrt}
\maketitle

\begin{abstract}         
Canonical quantization has served wonderfully for the quantization of a vast number of classical systems. That includes single classical variables, such as $p$ and $q$, and numerous classical Hamiltonians $H(p,q)$, as well as field theories,
 such as $\pi(x)$ and $ \phi(x)$, and many classical Hamiltonians $H(\pi,\phi)$. However, in all such systems there are situations for which canonical quantization fails. This includes certain particle and field theory problems. Affine quantization involves a simple recombination of classical variables that lead to a new chapter in the process of quantization, and which is able to solve a vast variety of normally insoluble systems, such as quartic interactions in scalar field theory in spacetime dimensions 4 and higher, as well as the quantization of Einstein's gravity in 4 spacetime dimensions. 
  \end{abstract}
  
  \section{Introduction}    Canonical 
  quantization is the leading quantization formulation, while there are other procedures that claim to be a substitute for canonical quantization. Affine quantization does not pretend to take the place of canonical quantization, but instead, it actually expands similar procedures of
  canonical quantization to solve problems that canonical quantization can not solve. For example,
  there are many models that are classically well behaved but that are unable to be quantized by canonical quantization. This article is focussed on how classical phase-space procedures can 
  point to an alternative set of basic quantum operators instead of the traditional basic quantum operators, namely, the momentum $P$ and the position $Q$, which satisfy $[Q, P]=i\hbar\, 1\!\!1$.
 In a sense, these operators are the foundation of canonical quantization. A new set of basic operators, $Q$ and
  $R$, which satisfy $[Q, R]=i\hbar\,Q$, and they work their magic that enables us to solve many problems,
  from simple particle models \cite{ME}, to quartic-interaction scalar fields in high spacetime dimensions such as $n\geq4$ 
  \cite{AA, FF}, to Einstein's general relativity in 4 spacetime dimensions \cite{ME, bqg, eq}.  
   
   While the references above offer fairly full stories, we will -- in keeping with the purpose of the
   present paper which is designed to offer a beginner-level  tutorial 
 on affine quantization --  present a basic outline of all three of the examples listed 
 above in this paper.

\section{Classical Variable Connections}
\subsection{Traditional classical variables} 
We first focus on a single degree of freedom. The familiar integral that leads to the equations of motion is derived from an action functional (with $p=p(t)$ and $q =q(t)\,, 0\leq t \leq T>0\,)$ given by
    \bn A =\tint_0^T [ p\,\dot{q} -H(p,q)]\, dt\;. \en 
 Standard stationary relations, including $\delta p(0)=0=\delta p(T)$ 
 and $\delta q(0)=0=\delta q(T)$, such that 
  \bn \delta A=\tint_0^T \{ (\dot{q}-\partial H(p,q)/\partial p\,)\,\delta p-(\dot{p}+\partial H(p,q)/
   \partial q
\,)\,\delta q\}\,  dt\,=0\;, \en
  leads to two, basic, equations of motion,
     \bn \dot{q}=\partial H(p,q)/\partial p\;\;\;,\hskip3em
     \dot{p}=-\partial H(p,q)/\partial q \;. \en
     
     These equations of motion apply whatever the range of the basic variables, $p$ and $q$.
     As examples of interest, let us assume (1) $-\infty<p,q<\infty$, which is conventional, or (2)
   $-\infty<p<\infty$ and $0<q<\infty$, which is less conventional, but very important (see below), and these, and other, ranges are generally chosen to be compatible with the equations of motion.
   
   \subsection{Traditional affine variables}
      New, and specific, phase-space coordinates may be introduced to Eq.~(1) by the relation
      $p\,dq = r \,ds $, and we   
      choose $ r\equiv p\,q$ and $s\equiv \ln(q) $. Note that the 
      choice of $s$ implies that
       $0<q<\infty$ and $-\infty<p<\infty$, while 
      $-\infty<r, s<\infty$.
      
      There are many pairs of variables that could be considered -- generally to offer help in solving the equations of motion -- but our choice of $r$ and $s$ is specifically chosen to offer a simple set of new variables that enjoys the domain $-\infty<r, s<\infty$.

      The new variables lead us to
      \bn A= \tint_0^T [r\dot{s}-H'(r,s)]\,dt \en  
      where $H'(r,s)\equiv H(p,q)$. We next introduce stationary relations
           \bn \delta A=\tint_0^T \{ (\dot{s}-\partial H'(r,s)/\partial r\,)\,\delta r-(\dot{r}+\partial H'(r,s)/
   \partial s
\,)\,\delta s \}\,  dt\,=0\;, \en 
where $\delta r(0)=0=\delta s(0)$ as well 
as $\delta r(T)=0=\delta s(T)$, which leads to two equations of motion given by
    \bn \dot{s}=\partial H'(r,s)/\partial r \;\;\;,\hskip 3em
         \dot{r}=-\partial H'(r,s)/\partial s\;. \en
    
 \section{Quantum Operator Connections}
 \subsection{Affine quantum operators}
 A conventional canonical quantization promotes classical variables to quantum operators such as
 $p\ra P$ and $q\ra Q$, and $r\ra R$ and $s\ra S$. These operators satisfy $[Q, P]=i\hbar 1\!\!1$
 and $[S, R]= i\hbar1\!\!1$ as well. But, since $s=\ln(q)$ or $q=e^s$, there should be some sort of
 connection between $Q$and $S$, such as $[Q,S]=0$, and  $S=\ln(Q)$ and, therefore, $Q=e^S$.
 We can support these equations by diagonalizing these variables. In particular, we have the states
 $|q\>$, $0<q<\infty$, where $Q|q\>=q|q\>$ and $\<q'|q\>=\delta(q'-q)$. Likewise, we have states $|s\>$, $-\infty<s<\infty$, 
 so that $S|s\>=s|s\>$ and $\<s'|s\>=\delta(s'-s)$. Finally, we observe that $Q|s\>=e^s|s\>$ and
 $S|q\>= \ln (q)|q\>$. Moreover, 
 \bn   && [S^n,  R]=i\hbar\,n\,S^{n-1}\no\\
        &&\sum_{n=1}^\infty (1/n!)\,[S^n, R]=i\hbar \sum_{n=1}^\infty (n/n!)\,S^{n-1} \no\\
        && [e^S-1, R]= i\hbar \,e^S\no \\
        && [e^S, R]=i\hbar\,e^S\no\\
        && [Q, R]=i\hbar\, Q\;, \label{xx} \en    
        When $0<Q<\infty$, its partner $P$ can not be self adjoint. However, for the same $Q$ a 
        substitute partner is $R$ which can be self adjoint along with $Q$.\footnote{Equation 
        \ref{xx} also applies if $-\infty<Q<0$ or, as a reducible case, $-\infty<Q\neq 0<\infty$. However,we will generally 
        focus on the case where $0<Q<\infty$.}

       This result helps point the way toward solubility. The quantum operators $P$ and $Q$, where $0<Q<\infty$,
 {\it guarantees that $P$ cannot be made self adjoint!} That situation implies that canonical quantization using $P$
 and $Q$ would not succeed. However, we can instead choose  $R$ and $Q$ {\it because they can both be self adjoint!} {\it While $p\ra P$ and $q\ra Q$ are the foundation of canonical quantization, $r\ra R$ and $q\ra Q$ are
 the foundation of affine quantization.}\footnote{In the author's previous work the classical
 term $r=p\,q$ was called $d=p\,q$ and the quantum operator $R=(PQ+QP)/2$ was called $D=(PQ+QP)/2$ ($d$ and $D$ was for `dilation'). For purposes of this article, we adopt the symbols $r=p\,q$ and $R=(PQ+QP)/2$, when
 $0<Q<\infty$. Hereafter, we will continue to focus mainly on $0<Q<\infty$.}
 
 \subsection{Coherent states}
 Coherent states provide a connection between the classical and the quantum realms. Simple coherent states rely on two classical variables and two quantum operators, the last of which are self-adjoint
 operators. For systems for which $-\infty<p, q<\infty$ and $Q$ and $P$ are self adjoint the canonical coherent states are given by
    \bn |p, q\>= e^{-iqP/\hbar}\,e^{ipQ/\hbar}\,|0\> \;, \en
    where $(\omega Q+iP)\,|0\>=0$. 
    
    If, instead,  $0<q<\infty$, then these coherent states fail
    and one may appeal to the affine coherent states.   
    The affine coherent states are given (where it is convenient to choose $q$ and $Q$ as dimensionless, and their usual dimension is absorbed by $p$ or $R$)
  \bn |p;q\>=e^{ipQ/\hbar}\,e^{-i\ln(q)\,R/\hbar}\,|\beta\>\;, \en
        with $[(Q-1)+iR/\beta]|\beta\>=0$.
        Here we introduce a semicolon in the coherent-state label to signify that affine items are under discussion, and not canonical items.
        
        \subsection{Quantum/classical transition}   
        \subsubsection{the canonical story}
        The quantum action functional leading to Schr\"odinger's equation is given, for normalized 
        Hilbert space vectors $|\psi(t)\>$, by
        \bn A = \tint_0^T \<\psi(t)| [i\hbar\,\pp/\pp t-\mathcal{H}(P,Q)]\,|\psi(t)\>\, dt\;.\en
        A general stationary variation leads to Schr\"odinger's equation,
          \bn i\hbar\,\pp |\psi(t)\>/\pp t=\mathcal{H}(P,Q)\,|\psi(t)\>\en
          and its adjoint.
          However, macroscopic observers can only vary a limited set of vectors, such as the coherent states. This leads to a reduced (r) action functional given by
      \bn A_r\hskip-1.3em&&=\tint_0^T\<p(t), q(t)|[i\hbar\,\pp/\pp t -\mathcal{H}(P,Q)]\,|p(t),q(t)\>\,dt \no \\
              &&=\tint_0^T[ p(t)\dot{q}(t) - H( p(t), q(t)]\,dt \;, \label{ss} \en
              and the final result appears to be a {\it classical action functional}, and, indeed, one      that has an `advantage' because $\hbar>0$ just as it is in the real world!
              
   To complete this story we examine the quantum Hamiltonian, $\mathcal{H}(P,Q)$, and the `enhanced' (because $\hbar>0$) classical Hamiltonian, $H(p,q)$,
          and we learn that
          \bn H(p,q) \hskip-1em &&=\<p.q|\mathcal{H}(P,Q)|p,q\> \no \\
           &&= \<0|\mathcal{H}(P+p, Q+q)|0\> \no \\
         &&  =\mathcal{H}(p,q)+\mathcal{O}(\hbar; p,q)\;. \en
          It follows that in the true classical limit, where $\hbar\ra 0$ and $H(p,q)\ra H_c(p,q)$ 
          ($c$ for classical),
          the quantum operators assume the positions of the classical variables in the Hamiltonian.
          
          Moreover, the favored phase-space coordinates that lead to a physically correct quantization, as Dirac observed \cite{dirac}, are Cartesian coordinates. This proposal is 
          confirmed  \cite{ME} by
          \bn d\sigma(p,q)^2\equiv 2\hbar\,[ |\!| \,d|p,q\>\,|\!|^2-|\<p,q|\,d|p,q\>|^2]=
          \omega^{-1}dp^2+\omega\, dq^2\;.\label{kk} \en
          These phase-space variables are the favored ones to promote to quantum operators to have
          a physically correct quantization.
          For a comparison purpose, which is coming up, we note that the two-dimensional flat space in (\ref{kk})
           is a  {\it constant zero curvature space}.
          
          The foregoing material was devoted to situations where $-\infty< p, q<\infty$. 
          We now turn our attention to situations where $0<q<\infty$ while $-\infty<p<\infty$; this analysis will be more streamlined.
          
          \subsubsection{the affine story}
         The quantum action functional for affine variables begins much as the canonical variable story, again for normalized Hilbert states $|\psi(t)\>$.
      In the new situation (with a prime to recognize $R$ in place of $P$) we have
        \bn A=\tint_0^T\<\psi(t)|[i\hbar\,\pp /\pp t-\mathcal{H}'(R,Q)\,]
        |\psi(t)\>\, dt\;, \en
        which, with stationary variations, leads directly to Schr\"odinger's equation and its adjoint. Restricting the variations to affine coherent states leads to
        \bn A_r \hskip-1.4em &&=\tint_0^T \<p(t);q(t)|[i\hbar\,\pp/\pp t-\mathcal{H}'(R, Q)]|p(t);q(t)\>\;dt \no\\
         &&=\tint_0^T [-q(t)\,\dot{p}(t)-H(p(t), q(t))]\;dt \;. \label{tt} \en
        Recalling that in this case both $q$ and $Q$ are dimensionless, it follows that
           \bn H(p,q)\hskip-1.3em&&=\<p;q|\mathcal{H}'(R, Q) |p;q\> \no\\
         &&  =\<\beta| \mathcal{H}'(R+ pqQ, qQ)|\beta\> \no \\
        &&     =\mathcal{H}'(pq,q)+\mathcal{O}(\hbar;p,q) \no \\
        &&  =\mathcal{H}(p,q)+\mathcal{O}(\hbar;p,q) \;. \label{ccc} \en
       Once again we see that $\mathcal{H}(p,q)=H_c(p,q)$ in the classical limit when $\hbar\ra0$.
       
       Are the favored phase-space variables Cartesian coordinates in the affine case? The answer is no. In particular, we find \cite{ME} that
       \bn d\sigma(p,q)^2\equiv 2\hbar [|\!|\,d|p;q\>\,|\!|^2-|\<p;q|\,d|p;q\>|^2]=
       \beta^{-1}q^2\,dp^2+\beta\,q^{-2}\,dq^2 \;.\en
       This two-dimensional space is not flat, but it is a {\it constant negative curvature space}, with a
       negative curvature that is $-2/\beta$. In the sense of having a constant curvature value
       this space is in the same category as the flat space. These phase-space variables are 
       the favored ones to promote to quantum operators in order to have a physically correct 
 quantization.\footnote{Unlike a flat plane, or a constant {\it positive} curvature surface (which holds the metric of three-dimensional spin coherent states), a space of constant negative curvature can not be visualized in a 3-dimensional flat space \cite{vv}. At every point in this space the negative curvature appears like a saddle having an `up curve' in the direction of the rider's chest and a `down curve' in the direction of the rider's legs.}

    \subsubsection{the classical/quantum story}
    Observe that the classical story, as derived from the proper quantum operators and the appropriate
    coherent states in (\ref{ss}) and (\ref{tt}), lead to a very similar classical story. So similar are their stories that
    if one of the quantizing procedures fails then there is a very good chance that the other
    quantization procedure could succeed. 
    
    This similarity enables us to claim that affine quantization should be considered as a parallel 
     road that starts in phase space and results in an acceptable quantum theory for selected problems, which is 
     exactly what canonical quantization has to offer for a different set of selected problems. 
     
     The union of canonical quantization and affine quantization is called `enhanced
     quantization', and this program can be found in \cite{eq}.
     
     Hereafter, we examine three classical
     systems that canonical quantization fails to solve but affine quantization is able to solve.
     These examples include a single degree-of-freedom model, and two field theories:
     one about quartic scalar fields in 4, and more, spacetime dimensions, and the other is Einstein's
     theory of gravity in general relativity in 4 spacetime dimensions.\footnote{Path integrals are commonly used to quantize classical theories as well. A mathematically sound procedure, which uses different sets of coherent states so as to feature specific quantizations, either a canonical quantization, an affine quantization, or a spin quantization, is discussed in \cite{eee},  Chap. 8.}
     
     \section{The ``Harmonic Oscillator''}
     The standard  harmonic oscillator entails $-\infty<p, q<\infty$ with a classical
     Hamiltonian $H(p,q)=(p^2+q^2)/2$. Its canonical quantization is well known and need not
     be repeated here. However, our example (which gives rise to the quotation marks in this section's
     title) has the same Hamiltonian, $H(p,q)=(p^2+q^2)/2$, with $-\infty<p<\infty$, but now
     $0<q<\infty$. For such a system we choose the classical Hamiltonian given by
       \bn H(p,q)=(p^2+q^2)/2=(pq\,q^{-2}\,pq+q^2)/2=(r\,q^{-2}\,r+q^2)/2=H'(r,q).\en
       According to (\ref{ccc}), we choose the quantum Hamiltonian to be
       \bn \mathcal{H}'(R,Q)=(R\,Q^{-2}\,R+Q^2)/2\;, \en
       and according to (\ref{xx}) we have $[Q, R]=i\hbar\,Q$. Adopting a Schr\"odinger 
       representation for $Q=x$, with $0<x<\infty$, it follows that
        $R=-i\half\hbar[x(\pp/\pp x)+(\pp/\pp x)x]$
        The appropriate Schr\"odinger equation is then given by
        \bn i\hbar\,\pp\,\psi(x,t)/\pp t \hskip-1em&&= \half \{-\quarter\hbar^2[x(\pp/\pp x)+(\pp/\pp x)x] \,x^{-2}\no\\     &&\hskip3em \times   [x(\pp/\pp x)+(\pp/\pp x)x]+x^2\}\;\psi(x,t) \no \\
             && =\half[-\hbar^2\pp^2/\pp x^2+\threequarter \hbar^2\,x^{-2}+ x^2]\,\psi(x,t) \;. \en
       
     This latter problem has also been examined using canonical quantization, an analysis which fails.
     That story can be found in  \cite{ME}, Sec.~1.5.
     
     \section{Scalar Quartic Interaction Field Theories \\in High Spacetime Dimensions}
     Quartic interactions for scalar field theories may have satisfactory canonical quantizations
     provided the spacetime dimension $n\leq 3$. For $n=4$ canonical quantization leads to 
     a free quantum theory, despite the coupling constant $g_o>0$ \cite{you}, and for $n\geq5$, the analysis becomes nonrenormalizable. 
     
     The classical version of such models is able to avoid high energy regions provided by the interaction term. For a finite overall energy the classical behavior of these models is well behaved. However, quantization can not avoid the domain issues that exist. Specifically, assume that
     the domain  for a free scalar field $\p(x)$ is $\mathcal{D}_{g_o=0} $, but when the interaction 
     is introduced, i.e., $g_o>0$, then the domain of the scalar field is immediately reduced, i.e.,
     $\mathcal{D}_{g_o>0}\subset\mathcal{D}_{g_o=0}$. The new, smaller domain survives even when
     $g_o\ra0$. This behavior is not significant for classical solutions, but such domain behavior 
     complicates the quantization of such models. 
     Since canonical quantization can not quantize them, let us try affine quantization. 
     
     It is newsworthy that these models can be solved via affine quantization, but that solution
     is {\it not} dependent on `positivity restrictions' for fields such as $0<\phi(x)<\infty$. 
     Instead, affine quantization leads to positive results simply by combining a momentum field  $\pi(x)$ and a scalar field $\p(x)\neq 0$, i.e., $\kappa(x)=\pi(x)\,\p(x)$, and using $\kappa(x)$ instead 
     of $\pi(x)$. The secret of affine quantization lies in being ready to accept 
     a new ground state dictated by the reduction of domains.\footnote{An affine
quantization for an ultralocal (meaning: no gradients) scalar model in \cite{bqg} 
     illustrates how the ground state for this model is not the usual free ground state. Such behavior is referred to as  `pseudofree'.}
     
       \subsection{Quantization of selected scalar field theories}
     In this section we examine a scalar field with a quartic interaction
     and we employ an affine quantization. The model has a standard classical Hamiltonian,
      \bn H(\pi,\phi)=\tint \{ \half[ \pi(x)^2+ 
      (\overrightarrow{\nabla}\phi(x))^2+m_o^2\,\phi(x)^2] +g_o\,\phi(x)^4\,\}\;d^sx\;;\en  
      here $s$ denotes the number of spatial coordinates.
       Next, we introduce the affine field $\kappa(x)\equiv \pi(x)\,\phi(x)$,
      and modify the classical Hamiltonian to become
       \bn &&H'(\kappa, \phi)=\tint \{ \half[ \kappa(x)\phi(x)^{-2}\kappa(x) +(\overrightarrow{\nabla}\phi(x))^2+m_o^2\,\phi(x)^2]\no \\
      &&\hskip12em +g_o\,\phi(x)^4\,\}\;d^sx\;.\en
      
      For quantization it follows that $\phi(x)\ra\hph(x)\neq0 $ plus $\kappa(x)\ra \hat{\kappa}(x)$,
      and we accept that $[\hph(x),\hat{\kappa}(y)]=i\hbar\,\delta(x-y)\,\hph(x)$. Additionally, we assume we have a Schr\"odinger representation in which $\hph(x)= \p(x)$ and
      \bn \hat{\kappa}(x)= -\half i\hbar [\p(x)(\delta/\delta\p(x))+
      (\delta/\delta\p(x)) \p(x)]\;.\en
      It follows that Schr\"odinger's equation is given by
       \bn &&\hskip-.5em i\hbar\,\pp\,\Psi(\p,t)/\pp t=
       \tint \big{\{} \half[\,\hat{\kappa}(x)\,\phi(x)^{-2}\,
       \hat{\kappa}( x)+ (\overrightarrow{\nabla}\phi(x))^2+m_o^2\,\phi(x)^2] \no \\
    &&\hskip14em  +g_o\,\p(x)^4\;\big{\}}\;d^sx\;    \Psi(\p,t) \:. \en
       It is noteworthy that $\hat{\kappa}(x)\, \p(x)^{-1/2}=0$, as well 
       as $\hat{\kappa}(x)\,\Pi_y\,\p(y)^{-1/2}=0$. 
       
       Solutions of this Schr\"odinger equation
       may be made simpler if its continuum formulation is first replaced by a suitable lattice regularization to facilitate finding solutions, and then, possibly, followed by an appropriate continuum limit; see \cite{bqg}. 
       
       There are some limited Monte Carlo studies of affine
       quantization of these scalar models; additional studies are welcome.

\section{Quantum Gravity}
     The basic variables for classical gravity are the symmetric-index metric $g_{ab}(x)$, $a, b, \cdots= 1, 2, 3$,
     and the matrix $ \{g_{ab}(x)\}>0$ which implies that $g(x)\equiv\det[g_{ab}(x)]>0$ as well. 
     In addition, the symmetric-index momentum is given by $\pi^{cd}(x)$, which are arbitrary real variables. 
     
     An important combination  is $\pi^a_b(x)\equiv \pi^{ac}(x)\,g_{bc}(x)$, with a summation over c. We name this combination term  `momentric' because it involves the {\it momen}tum and the me{\it tric}.
 
 To obtain the affine field we should combine the momentum and metric fields, which points us to the momentric field even though there is a difference in the number of independent metric terms (six) and momentric terms (nine). The Poisson brackets for these fields leads to
      \bn &&\{\pi^a_b(x),\pi^c_d(x')\}=
   \half\,\delta^3(x-x')\s[\delta^a_d\s \pi^c_b(x)-\delta^c_b\s \pi^a_d(x)\s]\;,    \no \\
       &&\hskip-.20em\{g_{ab}(x), \s \pi^c_d(x')\}= \half\,\delta^3(x-x')\s [\delta^c_a g_{bd}(x)+\delta^c_b g_{ad}(x)\s] \;,      \\
       &&\hskip-.30em\{g_{ab}(x),\s g_{cd}(x')\}=0 \;. \no  \en
        Unlike the canonical Poisson brackets, {these Poisson brackets suggest that they are equally   valid if $g_{ab}(x)\ra -g_{ab}(x)$, and thus there could be separate realizations of 
        each choice.\footnote{An earlier analysis of these issues appears in \cite{fff}.}
        
Passing to operator commutations,  we are led by (\ref{ccc}) and suitable coherent states \cite{ME},
to promote the Poisson brackets to the operators
 \bn   &&[\hp^a_b(x),\s \hp^c_d(x')]=i\s\half\,\hbar\,\delta^3(x-x')\s[\delta^a_d\s \hp^c_b(x)-\delta^c_b\s \hp^a_d(x)\s]\;,    \no \\
       &&\hskip-.10em[\hg_{ab}(x), \s \hp^c_d(x')]= i\s\half\,\hbar\,\delta^3(x-x')\s [\delta^c_a \hg_{bd}(x)+\delta^c_b \hg_{ad}(x)\s] \;, \\
       &&\hskip-.20em[\hg_{ab}(x),\s \hg_{cd}(x')] =0 \;. \no  \en
There are two irreducible representations of the metric tensor operator consistent with these
commutations: one where the matrix $\{\hg_{ab}(x)\}>0$, which we accept, and one where the matrix $\{\hg_{ab}(x)\}<0$, which we reject. 

   The classical Hamiltonian for our models is given \cite{adm} by
        \bn H(\pi, g)=\tint \{ g(x)^{-1/2} [\pi^a_b(x)\pi^b_a(x)-\half \pi^a_a(x)\pi^b_b(x)] 
           +g(x)^{1/2}\,^{(3)}\!R(x)\}\;d^3x, \en
 where $^{(3)}\!R(x)$ is the 3-dimensional Ricci scalar. For the quantum operators
we adopt a Schr\"odinger representation
 for the basic operators: specifically $ \hat{g}_{ab}(x)=g_{ab}(x)$ and 
   \bn \hat{\pi}^a_b(x)=-\half i \hbar\,[\,g_{bc}(x)\,(\delta/\delta\,g_{ac}(x))+(\delta/\delta\,
   g_{ac}(x)))\,g_{bc}(x)\,]\;.\en
 It follows that the Schr\"odinger equation is given by
                \bn && i\hbar\,\pp\;\Psi(\{g\}, t)/\pp t)=\tint \{ [\hat{\pi}^a_b(x)\, g(x)^{-1/2} \,\hat{\pi}^b_a(x)-\half \hat{\pi}^a_a(x)\, g(x)^{-1/2} \,\hat{\pi}^b_b(x)] \no\\
         &&\hskip10em  +g(x)^{1/2}\,^{(3)}\!R(x)\}\;d^3x\;\Psi(\{g\}, t)\;, \label{rrr} \en
         where $\{g\}$ represents the 
          $g_{ab}(x)$ matrix. 
          
          It is noteworthy \cite{bqg} {that
     \bn \hat{\pi}^a_b(x)\,g(x)^{-1/2}=0\;\;\:, \hskip2em \hat{\pi}^a_b(x)\,\Pi_y \,g(y)^{-1/2}=0\;.\en
          Indeed, these two equations imply that 
           the factor $g(x)^{-1/2}$ can be moved to the left in the Hamiltonian in 
           (\ref{rrr}). Using that fact we can change the Hamiltonian operator, essentially
           by multiplying the Hamiltonian by $g(x)^{1/2}$, and using that expression to make the    result a simpler approach to fulfill the Hamiltonian constraints \cite{adm}
           to seek Hilbert space states $\Omega(\{g\})$ such that
           \bn  \{ [\hat{\pi}^a_b(x)\,\hat{\pi}^b_a(x)-\half \hat{\pi}^a_a(x) \,
           \hat{\pi}^b_b(x)] 
           +g(x)\,^{(3)}\!R(x)\} \;\Omega(\{g\})=0\;.  \en
           
           The analysis this far deals with the most difficult hurdle
           in the quantization of gravity. Further studies in this story are available in \cite{ME} as well as \cite{bqg}.  
           
   \section{Overview and Completeness}
   Let us begin by returning to a single degree of freedom, like $p$ and $q$, in the connection of
   canonical and affine procedures. The common action functional for canonical procedures is
   given by
      \bn A=\tint_0^T \{ p\,\dot{q}-H(p, q)\,\}\;dt \;. \label{kkk} \en
  As we presented earlier, stationary variations lead to the classical equations of motion.
  Selection of special phase-space variables, namely the `Cartesian coordinates' \cite{ME}, say $p$ and $q$, should be promoted to quantum operators, namely $P$ and $Q$, which obey
  $[Q, P]=i\hbar 1\!\!1$. Reduced to these few sentences, we have
  captured the essence of canonical quantization. We now turn to affine quantization.
  
  We modify the classical action functional as follows by featuring a new variable $r\equiv  p\,q$ 
  in place of $p$. This change leads to
    \bn  A=\tint_0^T \{\, r\,\dot{q}/q-H'(r,q)\,\}\;dt \;.\en
    Stationary variations again lead to the relevant equations of motion.
    Observe that the term $\dot{q}/q$ can deal with $0<q<\infty$, $-\infty<q<0$, or
    $-\infty<q\neq0<\infty$, where the factor $1/q$ leads to the requirement that $q\neq0$.
      As shown in Sec.~3.3.2, the new coordinates are $r$ and $q$, and the usual Poisson bracket 
      becomes $\{q, r\}= q$. These variables are then promoted to operators $R$ and $Q$, with
      $ [Q, R]=i\hbar\,Q$, along which, we claim, they describe the basic operators of affine quantization.
      
      As the reader sees a new commutator in the last sentence, they may be worried that there could be an avalanche of `new, basic' commutators.  To test that notion, less us consider the
      commutation $[Q^m, R]=i\hbar\,m\,Q^m$. Ignoring the case of $m=0$, we can feature
      $[Q^m, R/m]=i\hbar\,Q^m$. This relation recovers the original one when we set $Q^m\ra Q'$
      and $R/m\ra R'$, which signals that no `new, basic' commutators arise from this exercise.
      Moreover, $[Q, R]=i\hbar\,Q$ leads to a constant negative curvature, the amount being
      dependent on the choice of the set of coherent states. As previously noted, canonical quantization leads to a two-dimensional surface which has a constant {\it zero} curvature for canonical quantization, a two-dimensional surface which has a constant {\it positive} curvature for a spherical surface, the radius of which is fixed by the dimension of the finite Hilbert space, 
      and a two-dimensional surface which has a constant {\it negative} curvature for
      affine quantization, the value of which is fixed by the fiducial vector for the coherent states. This set of constant, curved two-dimensional surfaces  completes
      the set of such items, meaning that this particular feature is complete. 
      
      This presentation,
      for a single degree of freedom, essentially scales to apply to standard field theories and Einstein's  gravity by dropping classical momentum fields in favor of classical affine fields.
      
      Affine quantization deserves to be accepted into the `select group of 
      successful quantization procedures' so that we can quantize more classical systems!

\end{document}